# Resonance in Putting

Robert D. Grober

Department of Applied Physics, Yale University

New Haven, CT  06520

Abstract

It is shown that the putting stroke of world class golfers can be described as the motion of a pendulum driven at twice its natural resonance frequency.  This model minimizes error in the speed of the putter head due to random errors in the magnitude of the applied forces, providing rational for why great players have developed this particular putting stroke.

Introduction

It has long been observed that the putting stroke of proficient putters resembles a pendulum. The mass is a combination of the club, arms, and shoulders/torso and the spring constant can be a combination of gravitational and biomechanical forces. However, it is not enough to know the biomechanics is that of a pendulum, one must also understand how to apply forces to drive the pendulum. In this paper a model is proposed for the application of force in the putting stroke. The model is based on observations of the tempo, rhythm, and timing of proficient putters. The model is also relevant for short chip shots, where the double pendulum aspect of the full golf swing is not yet developed.

Guidelines for the Model

There are three fundamental observations regarding the putting stroke of proficient golfers on which the model is based [1]. The first is the observation that the putter head is moving at constant speed as it impacts the ball. The second observation is that the total duration (i.e. length of time) of the putting stroke is relatively insensitive to the length of the putt (i.e. the intended initial velocity of the ball). The final observation is that the ratio of the duration of the backswing, $\tau_b$, to the duration of the downswing, $\tau_d$, in the putting stroke is close to two,. $\tau_b/\tau_d \approx 2$. A goal of this paper is to provide some insight as to why proficient golfers have evolved a putting stroke with these characteristics.

The Equation of Motion

This paper is based on the equation of motion for the simple harmonic oscillator,

$$m\ddot{x} + kx = f(t),  \qquad (1)$$

where $x$ is displacement, $m$ is mass, and $f$ is force. This equation of motion can be used to model the pendulum-like motion of the putting stroke if one defines $x$ as the angle of the pendulum, $m$ as the moment of inertia of the pendulum, $k$ as the restoring force provided by both gravity and biomechanics, and $f$ as the torque that drives the pendulum. The torque is generated by application of a force which acts perpendicular to the long axis of the pendulum and along the direction of motion. This is distinct from the centripetal force, which acts along the long axis of the pendulum. The distinction is important, as centripetal force will be discussed towards the end of the paper.

This equation of motion ignores all damping and nonlinearity. We assume nonlinearities to be negligible and discuss the effect of damping at the end of the paper. Throughout this paper, $\dot{x}$ denotes the velocity of $x$ (i.e. first derivative with respect to time) and $\ddot{x}$ denotes acceleration of $x$ (i.e. second derivative with respect to time).

The equation of motion is generally rewritten in the convenient notation

$$\ddot{x} + \omega_0^2 x = \frac{f(t)}{m}$$

where the resonant frequency $\omega_0^2 = k/m$. The initial conditions are $x(0) = 0$ and $\dot{x}(0) = 0$.

A Preliminary Force Model

The condition that the putter be moving a constant speed as it impacts the ball, i.e. $\ddot{x} = 0$ at $x = 0$, can only be accommodated if the applied force is zero as the putter approaches impact. The observation that the duration of the putting stroke, $\tau_b + \tau_d$, is independent of the length of the intended putt is accommodated by requiring the temporal evolution of the driving force to be related to the resonant frequency of the mechanical system. One simple way of applying force consistent with both of these requirements is to 1) initiate the backswing with an impulsive force; 2) allow the pendulum move freely until the transition between backswing and downswing; 3) apply a second impulsive force opposite in direction to the initial force at the beginning of the downswing; and then 4) let the pendulum move freely to impact. Fig. 1 shows how these forces might evolve in time. The red curve is the force profile that initiates the backswing. The blue curve is the force profile at the beginning of the downswing. As drawn, the force profiles are of equal magnitude, though this need not be the case.

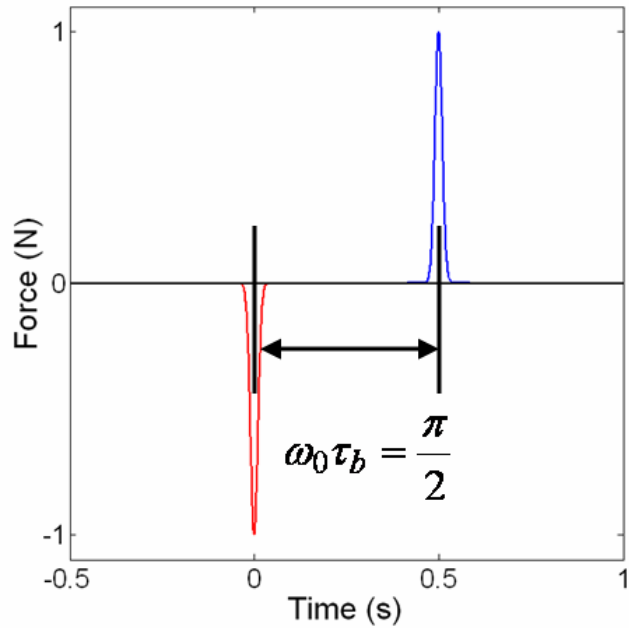

Fig. 1: Force profile for impulsive forces for the case $\omega_0 = \pi$. The red curve is the force profile that initiates the backswing. The blue curve is the force profile at the beginning of the downswing. As is discussed in the text, the time between the two impulses is defined by the resonant frequency of the pendulum and the force impulses are of equal magnitude.

Consider applying the force $f(t) = \Delta p\, \delta(t)$ at the beginning of the backswing, where $\delta(t)$ is a delta-function. The impulse of this force is the amount by which the momentum of the system is changed due to the application of the force, $\Delta p = \int f(t)\,dt$. We consider the case $\Delta p = -p_b$. The resulting motion of the pendulum is

$$x(t) = -\frac{p_b}{m\omega_0} \sin \omega_0 t$$

The backswing ends at the point $t = \tau_b$ such that $\omega_0 \tau_b = \frac{\pi}{2}$. At this point, the state of the system is $x(\tau_b) = -\frac{p_b}{m\omega_0}$ and $\dot{x}(\tau_b) = 0$.

At the top of the backswing, another impulsive force is applied but in the opposite direction to the initial force,

$$f(t) = p_d\, \delta(t - \tau_b)$$

The resulting motion of the pendulum is

$$x(t) = \frac{1}{m\omega_0}\left[-p_b \cos \omega_0 (t - \tau_b) + p_d \sin \omega_0 (t - \tau_b)\right]$$

The pendulum returns to the origin at the time $t = \tau_b + \tau_d$, where $\tau_d$ is the duration of the downswing and is determined by the condition

$$\tan \omega_0 \tau_d = \frac{p_b}{p_d}.$$

The velocity at the origin is then given as

$$\dot{x}(\tau_b + \tau_d) = \frac{\sqrt{p_b^2 + p_d^2}}{m}.$$

Note that the total time $\tau_b + \tau_d$ is independent of the final velocity as long as the ratio $\dfrac{p_b}{p_d}$ is held constant.

One would like to choose $p_b$ and $p_d$ consistent with the observation that the ratio of the duration of the backswing to the duration of the downswing is of order two, $\dfrac{\tau_b}{\tau_d} = 2$. We have already determined $\omega_0 \tau_b = \dfrac{\pi}{2}$. To achieve the observed backswing to downswing ratio, one must obtain $\omega_0 \tau_d = \dfrac{\pi}{4}$. This occurs for the condition $p_b = p_d$. Thus, *the tempo ratio has the observed value when the magnitude of the impulse at the beginning of the backswing equals the magnitude of the impulse at the beginning of the downswing.*

While it seems instinctive that the stroke is easier to manage if the forces involved are symmetric, one might ask if there is a deeper reason behind this result. Consider the momentum of the putter at impact

$$p_0 = m\dot{x}(\tau_b + \tau_d) = \sqrt{p_b^2 + p_d^2}$$

One can ask how this momentum varies if the impulsive forces have some random error. Taking a differential, one obtains

$$\frac{\Delta p_0}{p_0} = \frac{p_b^2}{p_0^2}\frac{\Delta p_b}{p_b} + \frac{p_d^2}{p_0^2}\frac{\Delta p_d}{p_d}$$

where this expression has been written in terms of the relative error $\Delta p/p$. If we assume the relative errors in the backswing and downswing forces are uncorrelated random variables with zero mean, $\langle \Delta p/p \rangle = 0$ and variance, $\langle (\Delta p/p)^2 \rangle = \sigma^2$, the statistics of the final momentum of the putter are

$$\langle \Delta p_0 \rangle = \frac{p_b^2}{p_0} \left\langle \frac{\Delta p_b}{p_b} \right\rangle + \frac{p_d^2}{p_0} \left\langle \frac{\Delta p_d}{p_d} \right\rangle = 0$$

and

$$\left\langle \left( \frac{\Delta p_0}{p_0} \right)^2 \right\rangle = \left( \frac{p_b^4 + p_d^4}{p_0^4} \right) \sigma^2$$

One can rewrite the variance in terms of a single parameter $p_b$, yielding

$$\left\langle \left( \frac{\Delta p_0}{p_0} \right)^2 \right\rangle = \left( \frac{p_b^4 + \left( p_0^2 - p_b^2 \right)^2}{p_0^4} \right) \sigma^2.$$

This function has a local minimum at $p_b = p_0/\sqrt{2}$, where it has the value

$$\left\langle \left( \frac{\Delta p_0}{p_0} \right)^2 \right\rangle = \frac{\sigma^2}{2}.$$

At the point of this local minimum, $p_b = p_0/\sqrt{2}$, one obtains $p_b = p_d$! *This analysis strongly suggests the use of equal forces at the beginning of the backswing and the beginning of the downswing serves to minimize the error of the speed of the putter due to random errors in the magnitude of the applied forces.*

In summary, the duration of the backswing is defined by the mechanical resonance, $\omega_0 \tau_b = \frac{\pi}{2}$. The duration of the downswing is defined by the ratio of the applied impulsive forces, $\tan \omega_0 \tau_d = \frac{p_b}{p_d}$. Thus, the tempo of the swing, which we define in terms of the values $\tau_b$ and $\tau_d$, is insensitive to the magnitude of forces so long as the ratio of the forces is held constant. The particular condition $p_b = p_d$ yields the tempo ratio observed for proficient putters, $\frac{\tau_b}{\tau_d} = 2$. This condition minimizes error in the speed of the putter due to random errors in the magnitude of the applied forces, which suggests a rationale for this particular putting stroke.

A major point in the above analysis is that the system must be driven resonantly. One obtains consistent tempo by using the mechanical resonance of the system as a natural clock. *This suggests a more quantitative description of what golfers describe as 'rhythm': a rhythmic golf stroke is the perception that the system is being driven resonantly.*

A More Realistic Force Model

It is highly unlikely that proficient putters impose delta-function forces on the putter. Any realistic force will have some finite duration. In the limit the impulsive forces are broadened sufficiently that the total force profile becomes continuous, a reasonable model of the force profile might be

$$f(t) = \begin{cases} -f_0 \sin 2\omega_0 t & 0 < \omega_0 t < \pi \\ 0 & \text{otherwise} \end{cases}.$$

This amounts to driving the putter at twice its natural resonant frequency. This force profile is shown schematically in Fig. 2. Note that the distance between the peaks of this force profile is the same as that for the impulsive force profile.

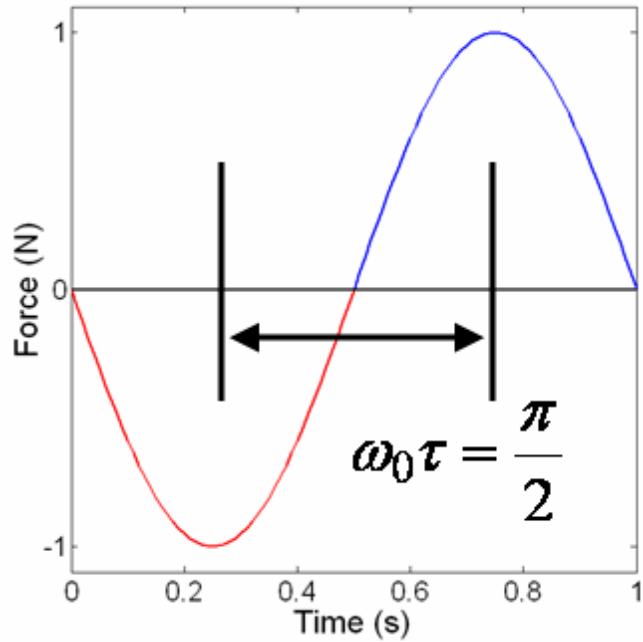

Figure 2: A force profile broadened from the impulsive force of Fig. 1 for the case $\omega_0 = \pi$. The distance between peak positions remains the same as for the impulsive force profile. The red region of the curve indicates a force that accelerates the club backwards and the blue region is a force that accelerates the club towards the ball.

Keeping the same notation as in the above analysis, it is convenient to write the force in units of momentum,

$$f(0 < \omega_0 t < \pi) = -\frac{3 p_0 \omega_0}{4} \sin 2\omega_0 t .$$

This particular normalization is chosen such that the momentum at impact is $p_0$. The solution for the resulting displacement is straightforward,

$$x(0 < \omega_0 t < \pi) = -\frac{p_0}{4 m \omega_0}\left(2 \sin \omega_0 t - \sin 2\omega_0 t\right).$$

This solution is shown in Fig. 3, where the curve has been color coded to show the time during which a backward going force is applied (red) when a forward going force is applied (blue).

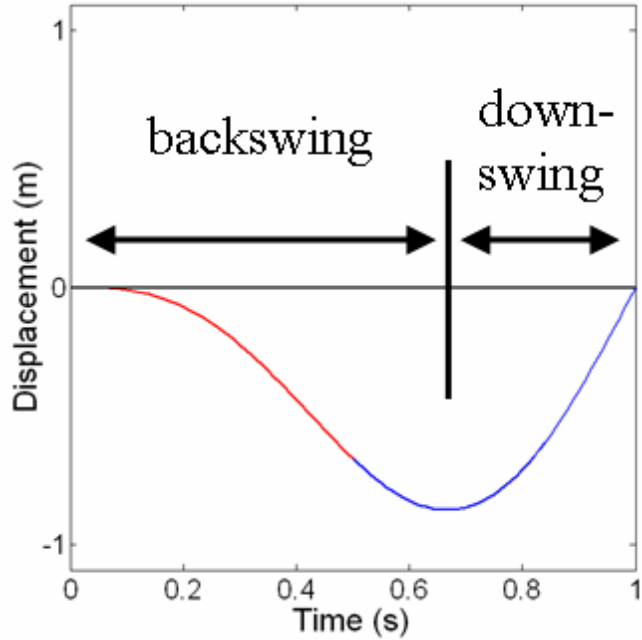

Fig. 3: Displacement as a function of time for the case $\omega_0 = \pi$. The curve has been color coded to show the time during which a backward going force is applied (red) and when a forward going force is applied (blue). Note that the total duration of the backswing is 2/3 sec while the duration of the downswing is 1/3 sec. Also note that the downward force (blue) is first applied while the club is still moving backwards.

All three required aspects of the model are obtained. First, the duration of the backswing is $\omega_0 \tau_b = \dfrac{2\pi}{3}$. The putter returns to the origin at $\omega_0(\tau_b + \tau_d) = \pi$.

Therefore, the tempo ratio is $\dfrac{\tau_b}{\tau_d} = 2$. Second, the velocity is constant as the club returns to the origin, i.e. $\ddot{x}(\omega_0 t = \pi) = 0$. Third, both $\tau_b$ and $\tau_d$ are insensitive to the magnitude of the applied force, thus the duration of the stroke is independent of the length of the putt.

A very interesting aspect of this solution is that the forward force is first applied while the club is moving backwards. This provides a very convenient way for the golfer to derive a cue reinforcing the direction in which to apply the forward going force: simply push against the momentum of the putter as it finishes the backswing.

Comparison of the Model with Data

In the following two figures, this simple model is compared to data taken on the putting stroke of a professional golfer whose putting stroke is representative of those with the properties described above. The data were taken using the Science and Motion measurement apparatus [2].

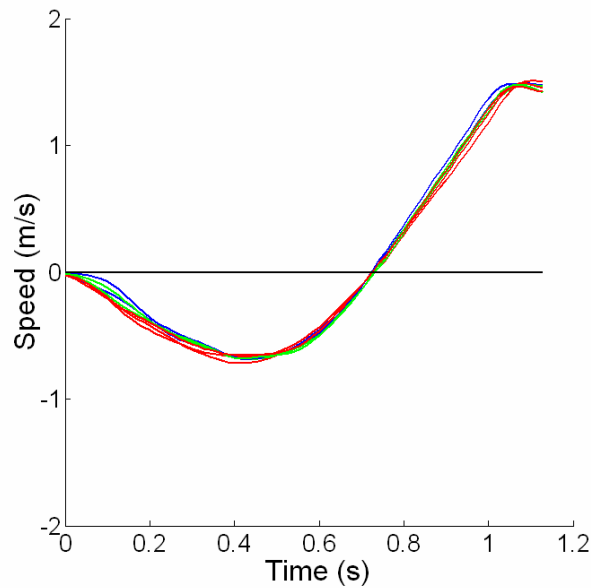

Figure 4: Velocity of the putter as a function of time for seven putting strokes made by a professional golfer whose stroke is representative of putting strokes with the properties described above. The data are all aligned relative to the point of transition between the backswing and downswing (i.e. the zero of the velocity profile in the vicinity of 0.7 seconds) and are normalized such that the integral of the absolute value of the velocity is equal for all seven data sets.

The data in Fig. 4 are the velocity of the putter as a function of time for seven strokes of the putter. The data are all aligned relative to the point of transition between the backswing and downswing (i.e. the zero of the velocity profile in the vicinity of 0.7 seconds) and are normalized such that the integral of the absolute value of the velocity is equal for all seven data sets.

The average of the data in Fig. 4 is fit to the model described above. The result of this fit is shown in Fig. 5. The average of the seven putting strokes is indicated by error bars, the width of which is the standard deviation of the seven data sets at each point in time. The data is fit to the model

$$\dot{x}(0 < \omega_0 t < \pi) = -A(\cos \omega_0 t - \cos 2\omega_0 t).$$

The frequency, $\omega_0$, is determined as $\omega_0 = \dfrac{2\pi}{3\tau_b}$, where $\tau_b$ is the average length of the backswing for the seven putting strokes. The amplitude, $A$, is determined by normalizing the model to the area under the velocity profile of the backswing for the average of the seven data sets. The resulting fit is shown as the solid red/blue line in Fig. 5. Consistent with the previous notation, the curve is color coded such that the red region corresponds to when the force is applied backward and the blue region is when the force is applied forward. For a relatively simple theory, it does a very respectable job of representing the data.

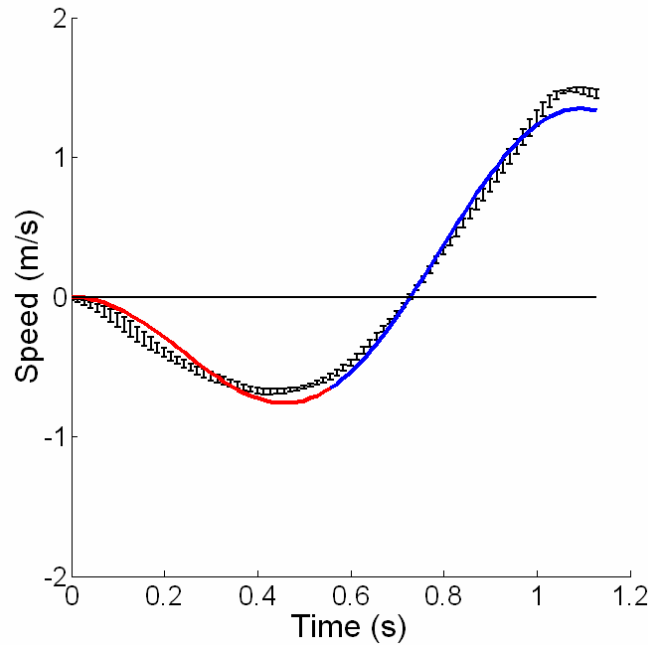

Figure 5: Fit of the model to the average of the data. The data is indicated by error bars, the width of which is the standard deviation of the seven data sets at each point in time. The model is shown as the red/blue line, where the red corresponds to when the force is applied backward and the blue is when the force is applied forward. The details of the fitting procedure are described in the text. For a very relatively simple theory, it does a respectable job of representing the data.

Force – Displacement Graphic

Given that this theory provides a relatively good fit to the data, as was shown in Fig. 5, it is interesting to ask if one can use the data to estimate the forces applied by the golfer. One can use the data to calculate the acceleration by differentiating the velocity profile. One can calculate the displacement by integrating the velocity profile. The resonant frequency was approximated by measuring the total duration of the backswing of the putting stroke, as described above. Given this input, one can estimate the resulting force profile using the equation of motion

$$\frac{f(t)}{m} = \ddot{x} + \omega_0^2 x.$$

Note that this gives force normalized by the mass. The result is shown in Fig. 6 as the dots connected by the solid black line. Note that this force profile is only as good as the validity of the equation of motion. The wiggles in the curve are likely not real, but rather are manifestations of shortcomings in the model. This curve is compared with the force profile corresponding to the best fit to the data, shown again with the red-blue color coding.

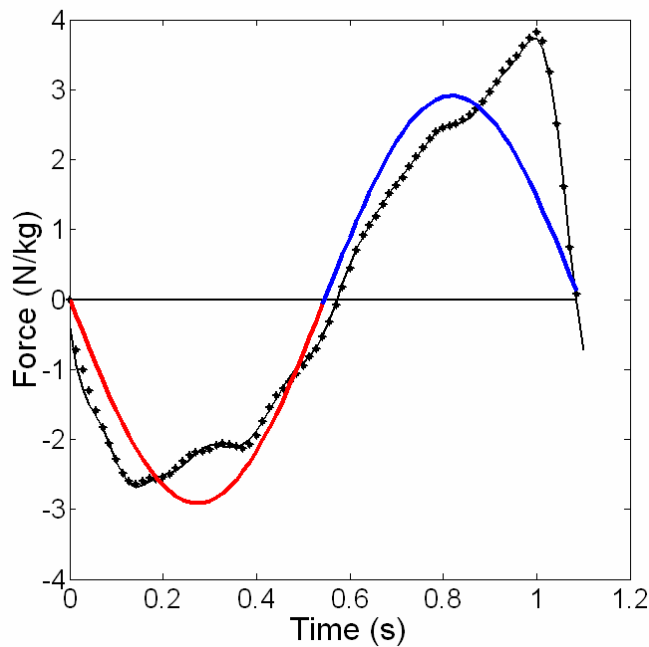

Fig. 6: The calculated force profile obtained from the velocity data and the equation of motion, as described in the text, compared with the model force profile used to obtain the best fit to the velocity profile data. The calculated force is displayed as black circles connected by a black line. The model force profile is indicated by the red-blue line. The red indicates the region where the applied force is oriented backward and the blue region is where the applied force is forward going.

Perhaps a more interesting way to view the data is in terms of a force-displacement graph, shown in Fig. 7. The calculated force profile for the averaged putting stroke is shown as the black dots connected by the solid black line. The fitted force profile is shown as the color coded line. This particular method of displaying the data is useful as it emphasizes where in the putting stroke the model force differs from the force derived from the data. As has been noted previously, the force model is pretty good at defining the overall shape of the curve. The largest differences occur during the last phase of the stroke, just before impact. This accounts for the fact that the velocity profile of the data has a slightly larger impact velocity than does the model, as can be seen in Fig. 5.

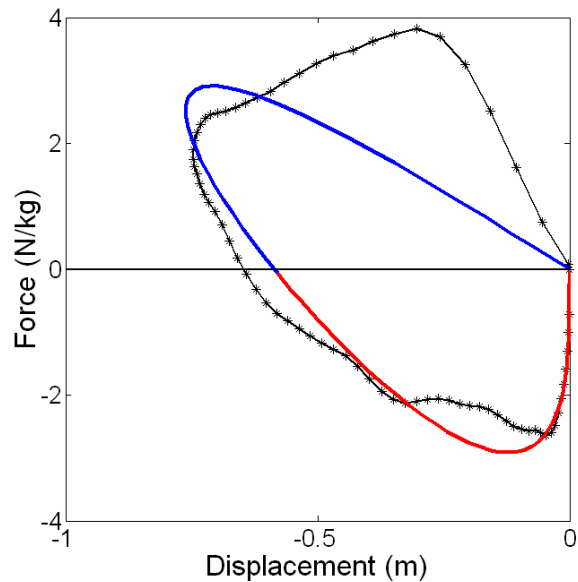

Fig. 7: A force displacement diagram, showing the applied force as a function of the position of the putter. The calculated force is displayed as black circles connected by a black line. The model force profile is indicated by the red-blue line, where the red region is where the force is applied backward and the blue region is where the force is applied forward. While the model force profile is pretty good at defining the overall shape of the curve, the two force profiles differ substantially in the last phase of the putting stroke, just before impact.

One might be concerned that the difference in force profiles between the model and data in this last phase of the putting stroke are sufficiently large as to call into question the validity of the model. However, as is seen in Fig. 5, the difference in velocity profile is relatively small. As is discussed in the next session, the velocity profile is relatively insensitive to changes in the force profile, so long as the force profile remains rooted in the second harmonic of the resonance.

Robustness of the Velocity Profile

An advantage of driving the putter at twice the resonant frequency is that the velocity profile is relatively insensitive to both the shape of the force profile and to damping.

The effects of perturbations in the force profile are best understood in terms of a Fourier analysis. If the force profile $f(t)$ contains a Fourier component of magnitude $F(\omega)$, the resulting influence on the analogous Fourier component of velocity scales as $\dot{X}(\omega) \propto \dfrac{\omega}{\omega_0^2 - \omega^2} F(\omega)$. Given a generalized force profile $f(t) \sim \sin 2\omega_0 t$, perturbations that preserve the timing introduce harmonics at integer multiples of the fundamental driving frequency; $4\omega_0$, $6\omega_0$, $8\omega_0$, $10\omega_0$, etc. Furthermore, if the perturbation maintains the symmetry (*i.e.* same force on the backswing as on downswing), then the harmonics are limited to odd integer multiples of the driving frequency; $6\omega_0$, $10\omega_0$, etc. The influence of these higher frequency force components on the velocity is characterized as

$$\frac{\dot{X}(n2\omega_0)}{\dot{X}(2\omega_0)} = \frac{3n}{4n^2 - 1} \frac{F(n2\omega_0)}{F(2\omega_0)}.$$

For the case $n=3$, the first term in a symmetric perturbation, the relative perturbation on the velocity profile is reduced by the factor 9/35 ~ 26% relative to the force profile. Therefore, perturbations that preserve both the timing and the symmetry of the force profile have a small effect on the velocity profile.

As examples of this principle, consider the three different applied force profiles, shown in Fig. 8(a). The black curve is the pure sine wave force profile; the red curve is a trapezoidal force profile; and the blue curve is an asymmetric force profile obtained by adding a forcing term at frequency $4\omega_0$ to the pure sine wave at frequency $2\omega_0$. The magnitudes of the forces have been scaled to yield the same velocity at impact. The resulting velocity profiles are shown in Fig. 3(b). Note that while there are significant differences in the force profiles, the resulting velocity profiles are very similar, consistent with the Fourier analysis described above. *This explains why proficient putters generate velocity profiles with very comparable attributes: the velocity profile is relatively insensitive to the details of the force profile as long as the force profile is resonant.*

All real systems exhibit damping, and one can reasonably expect the putting stroke is no exception. Damping is modeled as a velocity dependent term in the differential equation and is parameterized in terms of the $Q$ of the system,

$$\ddot{x} + \frac{\omega_0}{Q}\dot{x} + \omega_0^2 x = \frac{f(t)}{m}$$

The effect of damping can also be understood in terms of a Fourier analysis. Following the analysis above, if the force profile $f(t)$ contains a Fourier component of magnitude $F(\omega)$, the resulting influence on the analogous Fourier component of velocity scales as

$$\dot{X}(\omega) \propto \frac{\omega}{\omega_0^2 - \omega^2 - i\dfrac{\omega_0}{Q}\omega} F(\omega)$$

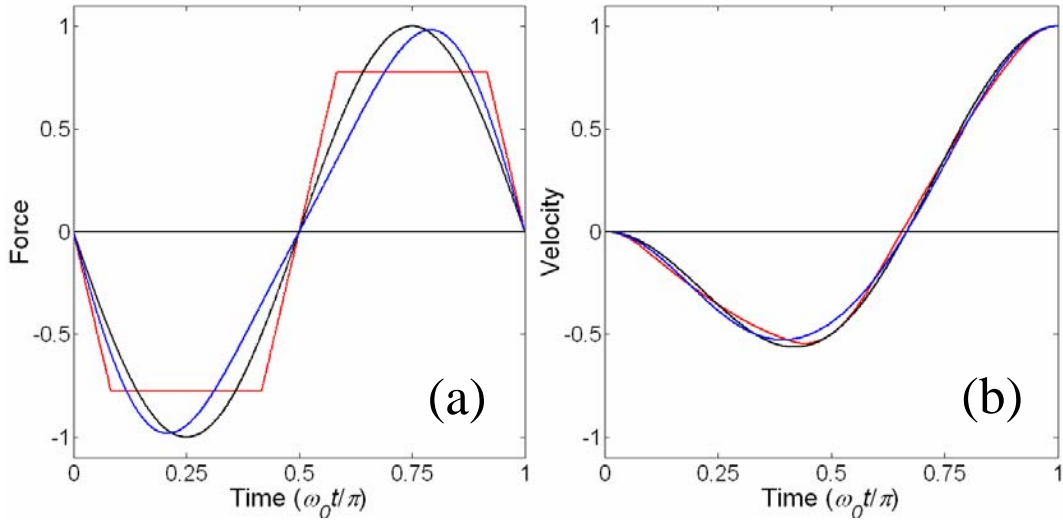

Figure 8: The graphic in (a) shows three different force profiles: a pure sine wave of frequency $2\omega_0$ (black), the same sine wave with the addition of a small forcing term at frequency $4\omega_0$ (blue), and a trapezoidal force profile (red). The graphic in (b) shows the resulting velocity profiles. The force profiles in (a) are normalized so as to yield the same velocity at impact in (b). As discussed in the text, the velocity profile is relatively insensitive to perturbations in the shape of the force profile, as long as the perturbations preserve the timing and the symmetry.

For systems driven at resonance, $\omega = \omega_0$, damping dominates the dynamics as it is the only surviving term in the denominator. However, for systems driven far from resonance, damping becomes less important, as is the case when driving the system at twice the resonant frequency.

We have modeled the effect of damping on the velocity profile for the driving force relevant to this discussion, $f(0 < \omega_0 t < \pi) \propto \sin 2\omega_0 t$. The results are shown in Fig. 9 for values of $Q$ ranging from two to twenty. As is shown in Fig. 9(a), the primary effect of damping on the system is to reduce the magnitude of the velocity and to reduce the total time of the putting stroke. However, the shape of the overall velocity profile is well preserved. This is shown in Fig 9(b), where time has been normalized for each stroke such that the duration of all putting strokes are equal and the magnitude of the velocity profiles have been normalized such that the velocities at impact are equal. Even for values of $Q$ as small as two, the primary aspects of the solution are mostly maintained: 1) the velocity is very nearly constant as the club approaches impact, 2) the tempo ratio is two, and 3) the duration of the putting stroke is independent of the magnitude of the applied force. This result is consistent with the expectation that damping plays a relatively small role for a system driven well above resonance.

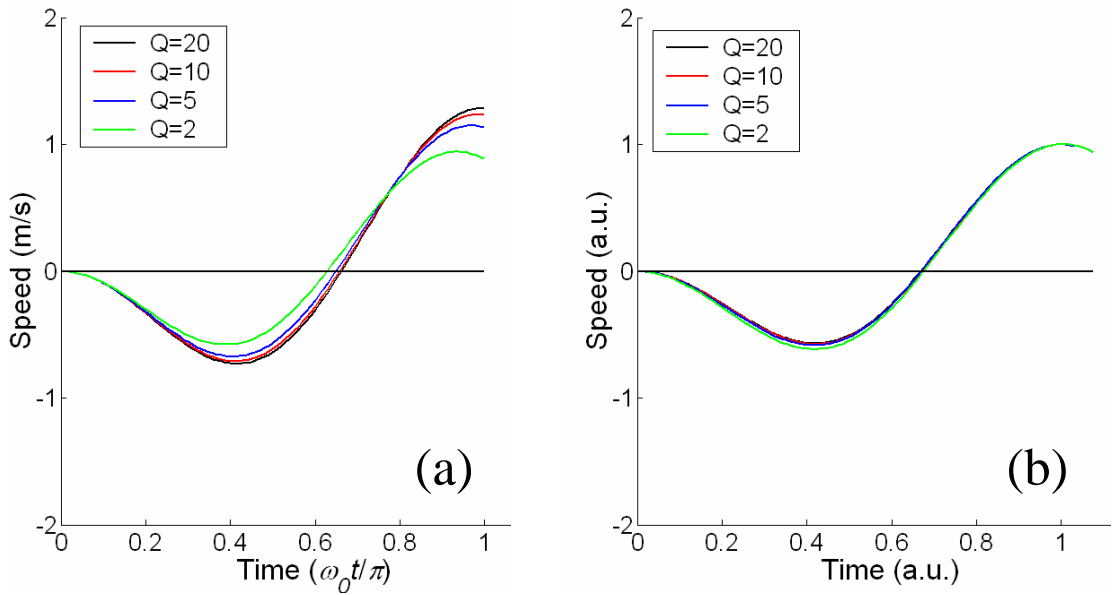

Figure 9: The graphic (a) is velocity as a function of time for values of $Q = 20, 10, 5,$ and 2 subject to the driving force $f(t) \propto \sin 2\omega_0 t$. Note that as the value of $Q$ decreases, the overall velocity decreases and the total time of the putting stroke decreases. The graphic (b) shows the same curves with the time normalized such that the duration of the putting strokes are equal and with the magnitude of the velocity normalized such that the velocities at impact are equal. Graphic (b) reveals that even for $Q$ as small as two, the effect of $Q$ on the overall shape of the velocity profile is modest.

Centripetal Force Considerations

The force described in the above analysis results in a torque that rotates the pendulum around its hub. The model requires the force go to zero as the putter approaches impact. The analysis did not consider the force opposing gravity that the golfer must apply to keep the putter from falling to the ground and the centripetal force the golfer must apply to keep the pendulum moving in a circle. The magnitude of this centripetal force is $\frac{mv^2}{R}$, where $m$ is the mass of the putter head, $v$ is the linear velocity of the putter head, and $R$ is the length of the pendulum. This centripetal force is not negligible. If $R \sim 1$m, then the centripetal force becomes comparable in magnitude to the gravitational force when $\dot{x} \sim 3$ m/s. For reference, this is approximately the putter head speed for a 40 foot putt on a reasonably fast green.

It is possible the centripetal force is a useful indicator for proficient golfers. Note the centripetal force is proportional to the square of the club head velocity, where as the torques involved in moving the putter require forces that are merely proportional to velocity. *Thus, on longer putts where the centripetal force is noticeable, the centripetal force near to impact may be a particularly sensitive indicator of club head speed.* Of course, one only becomes aware of these centripetal forces if the torques involved in moving the putter go to zero in the vicinity of impact. If the golfer aggressively accelerates the putter through impact, the centripetal forces are much less noticeable. Also, if one uses a "belly" putter, where the shaft of the putter is wedged against the belly of the golfer, it is unlikely the golfer will be aware of these centripetal forces.

A Hypothesis Regarding Short Putts

The above analysis suggests the putting stroke of proficient putters has evolved to minimize error in the velocity of the putter head at impact. A requirement of this analysis is that the putter head be moving with constant velocity at impact, which requires the applied force be zero at impact. This means the golfer exerts little physical control over the putter head at impact. This is a reasonable strategy for long putts, where the premium is on leaving the ball near to the hole. However, one might expect the strategy to shift as the length of the putt becomes shorter. In these situations, the length of the putt is less important relative to the direction of the putt. One might expect for these shorter putts that a better strategy would be to apply force to the putter head through impact, thereby enabling more control over the orientation of the putter head when it impacts the ball. One approach would be to use the same basic force profile described above, modified to allow a finite force at the very end of the putting stroke. Such a putting stroke would be characterized by a tempo ratio $\frac{\tau_b}{\tau_d} > 2$ and a modestly accelerating velocity at impact, but would preserve the rhythmic nature of the stroke. In terms of the graphic in Fig. 7, this putting stroke would deviate even more significantly from the model force profile in the very final phase of the putting stroke.

As a quantitative example of such a stroke, consider the following. The force profile discussed in the work above assumes $f(t) \propto \sin 2\omega_0 t$. One can represent $\sin 2\omega_0 t$ as the product function, $\sin 2\omega_0 t = 2\cos \omega_0 t \sin \omega_0 t$, which can be thought of as a driving term, $\cos \omega_0 t$, modulated by an envelope term, $\sin \omega_0 t$. One obtains the short putt force profile hypothesized above by modifying the envelope term such that

$f(t) \propto \cos\omega_0 t \sin(\omega_0 - \Delta)t$, where $\Delta$ is small relative to $\omega_0$. This type of force profile would preserve the resonant nature of the putting stroke but allow for a small forward force at impact, allowing the golfer more control over the club face.

Summary

It has been shown that the putting stroke of world class golfers can be described by a model in which a pendulum is driven at twice its natural resonance frequency. This model provides 1) a constant putter head speed at impact, 2) a putting stroke of total duration that is insensitive to the intended speed at impact and 3) a tempo ratio in which the backswing is of twice the duration as the downswing. This model happens to minimize error in the speed of the putter head due to random errors in the magnitude of the applied forces, providing rational for why great players have developed this particular putting stroke.

That the forcing term is at twice the resonant frequency of the pendulum renders the resulting velocity profile particularly insensitive to the exact shape of the force profile, so long as the force profile remains rooted in the second harmonic of the resonance. This suggests how it is possible different golfers may apply modestly differently force profiles but yield very comparable velocity profiles.

There are a couple of lessons one can take from this analysis:

1) This model serves to emphasize the importance of the forces applied during the backswing. The force applied in the backswing should be about equal to the force applied in the downswing. Within the context of this model, the length of the backswing is proportional to the speed of the club at impact. It is common counsel for golf pros to teach that the length of the backswing increases as the length of the putt increases.

2) The force transitions from backward to forward before the club transitions from backswing to downswing. This is very important, as it provides the golfer with a sense of direction. One can sense the direction of the downswing by pushing against the inertia of the moving club as it completes the backswing. Thus the transition is a dynamic process, providing a cue for the direction from the backswing.

3) The nice thing about $\omega_0(\tau_b + \tau_d) = \pi$ is that one can get a feel for this tempo by continuously and repeatedly swinging the club back and forth at resonance, in exactly the same manner one would swing a pendulum. The duration of the actual stroke is exactly the amount of time it takes for this pendulum like motion to swing the putter half a cycle (i.e. from the address position moving backward, to the address position moving forward). In fact, one often observes golfers instinctively doing this before they hit a putt.

4) This analysis suggests some quantitative definitions of tempo, timing, and rhythm in the putting stroke. Tempo is the pace of the swing, measured by $\tau_b$ and $\tau_d$. Timing is the process by which one returns the club to the ball with the putter head moving at constant velocity. Rhythm is how the club moves between the endpoints. As is suggested in this paper, a putting stroke is perceived as being "rhythmic" when is driven by forces that are harmonically related to its natural resonance.

Acknowledgments:

The author has benefited from conversations with Marius Filmalter, Hank Haney, Christian Marquardt, John Novosel, Peter Pulaski, Howard Twitty, and Grant Waite.